\documentstyle[11pt,newpasp,twoside,epsf]{article}
\def\edcomment#1{\iffalse\marginpar{\raggedright\sl#1\/}\else\relax\fi}
\reversemarginpar

\begin{document}
\title{Spectroscopy and Doppler tomography of BZ Ursae Majoris in quiescence} 

\author{V.V. Neustroev$^1$, A. Medvedev$^2$, S. Turbin$^3$, N.V. Borisov$^4$}
\affil{
$^1$ Department of Astronomy and Mechanics, Udmurtia State University,
Izhevsk, 426034, Russia\\
$^2$ Moscow State University, Moscow, Russia\\
$^3$ Izhevsk State Technical University, Izhevsk, Russia\\
$^4$ Special Astrophysical Observatory, Nizhnij Arkhyz, Russia
}

\section{Introduction}

BZ UMa is a little-studied, very infrequently outbursting dwarf nova with mean 
intervals longer than 5.2 yr, whose orbital period has been determined as 97.8 
minutes (Ringwald, Thorstensen, \& Hamway 1994).

BZ UMa was observed on January 08, 1995 using the SP-124 spectrograph of the 
6-m telescope of the Special Astrophysical Observatory. A total of 39 spectra were 
taken in the wavelength range 3950-5000 {\AA} with a dispersion of 1.1 {\AA}/channel. Individual exposure times were 200-300 s, the total duration of the observations was 
about 4.5 hours. 

The mean spectrum of BZ UMa at first sight is typical for cataclysmic variables 
(Fig. 1, top panel). 
However it is possible to detect some peculiarities. First of all we note 
unusual strong Balmer lines: we found equivalent widths and relative intensities 
for H$_{\beta }$ of more than 150 {\AA} and 7.5 respectively. Secondly, BZ UMa 
shows the triple-peaked lines profiles, which consist of the usual double-peaked 
profiles from an accretion disk, and of additional narrow component at the lines 
center (Fig. 1, bottom panel). The visibility of the latter improves from 
H$_{\beta }$ to H$_{\epsilon }$. The origin of this component is unclear. Key for understanding of nature of the emission source of the spike is the study of its 
radial velocities.

\section{Results}

First of all, we have determined the radial velocity semi-amplitude of the white 
dwarf and ${\gamma }$-velocity of the system. 
Our best fit parameters are 
$\gamma$=-73 $\pm$ 13 km s$^{-1}$ and K$_{1}$=57 $\pm$ 2 km s$^{-1}$.
Velocity parameters of the central spike measured after fitting the profile of the 
H$_{\delta }$ line with multiple gaussian functions are following: 
$\gamma$=41 $\pm$ 12 km s$^{-1}$ and K=79 $\pm$ 18 km s$^{-1}$. 
It can be seen, that the semi-amplitude K is consistent with the measurement 
based upon the whole emission lines and their wings, while $\gamma$-velocity 
of the central spike is much more.

The spike at the center of all the lines is an unusual puzzling feature of BZ UMa. 
At present the only one binary is known, namely double-degenerate binary GP Com, 
whose spectra demonstrate the like triple-peaked profiles of the emission lines. 
In GP Com the central spike has been ascribed to emission from a nebula or the 
accretion white dwarf (Marsh 1999). Really, any other part of the binary moves 
too fast or would form too broad a profile. 
Our measurements support the accretor origin for the central spike.
Nevertheless, both the accretor and the nebula hypotheses have complexities, 
which Marsh in detail considers. 

Now we can make a conclusion, that GP Com is the nearest equivalent of BZ UMa. 
Of course, GP Com is binary, which both members are degenerate, whereas BZ UMa 
is classical dwarf nova where a secondary is the normal M5.5 dwarf star 
(Ringwald et al. 1994). However both systems have a number of the unusual puzzling 
but similar observation properties. New more long and qualitative observations 
are extremely necessary for their explanation.

\begin{figure}
\plotone{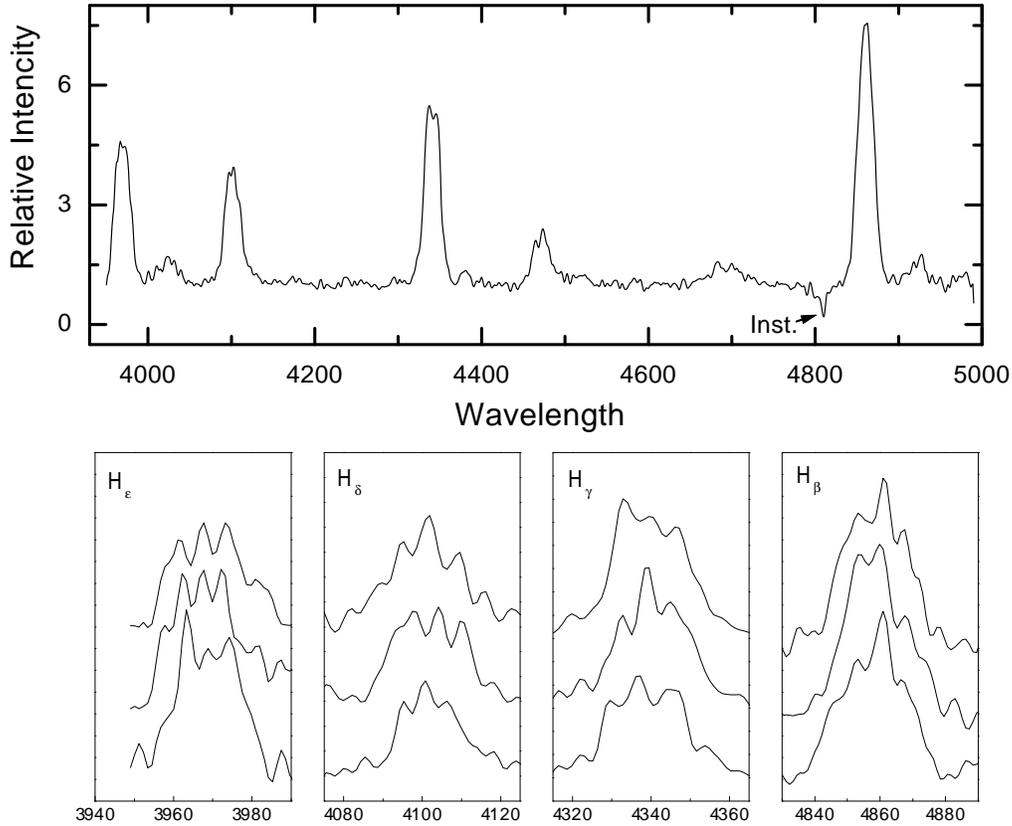}
\caption{The average normalized spectrum (top panel) and examples of the 
triple-peaked profiles of the major emission lines (bottom panel) of BZ UMa}
\end{figure}

\end{document}